\documentclass{midl} 

\usepackage{mwe} 

\usepackage{hyperref} 

\usepackage{array} 
\usepackage{graphicx,multirow}

\usepackage{bbm} 

\jmlrvolume{-- Under Review}
\jmlryear{2022}
\jmlrworkshop{Full Paper -- MIDL 2022 submission}
\editors{Under Review for MIDL 2022}

\title[Weakly Supervised Contrastive Learning]{Weakly Supervised Contrastive Learning for Better Severity Scoring of Lung Ultrasound}

\midlauthor{\Name{Gautam Rajendrakumar Gare\nametag{$^{1}$}} \Email{gautam.r.gare@gmail.com}\\
\Name{Hai V. Tran\nametag{$^{2}$}} \\
\Name{Bennett P deBoisblanc\nametag{$^{2}$}} \\
\Name{Ricardo Luis Rodriguez\nametag{$^{3}$}} \\
\Name{John Michael Galeotti\nametag{$^{1}$}} \\
\addr $^{1}$ Robotics Institute and Dept. of ECE, Carnegie Mellon University, Pittsburgh, USA \\
\addr $^{2}$ Dept. of Pulmonary and Critical Care Medicine, Louisiana State University, New Orleans, USA \\
\addr $^{3}$ Cosmeticsurg.net, LLC, Baltimore, USA
}

\begin{document}

\maketitle

\begin{abstract}
With the onset of the COVID-19 pandemic, ultrasound has emerged as an effective tool for bedside monitoring of patients. Due to this, a large amount of lung ultrasound scans have been made available which can be used for AI based diagnosis and analysis. Several AI-based patient severity scoring models have been proposed that rely on scoring the appearance of the ultrasound scans. AI models are trained using ultrasound-appearance severity scores that are manually labeled based on standardized visual features. We address the challenge of labeling every ultrasound frame in the video clips. Our contrastive learning method treats the video clip severity labels as noisy weak severity labels for individual frames, thus requiring only video-level labels. We show that it performs better than the conventional cross-entropy loss based training. We combine frame severity predictions to come up with video severity predictions and show that the frame based model achieves comparable performance to a video based TSM model, on a large dataset combining public and private sources.
\end{abstract}


\section{Introduction}

Lung Ultrasound (LUS) imaging has presented itself to be an effective bedside tool for monitoring COVID-19 patients \cite{Mento2020On2019, Raheja2019ApplicationReview, Amatya2018DiagnosticSetting}. Several AI based applications have emerged that help with diagnosis and identification of COVID-19 lung biomarkers \cite{Born2021AcceleratingAnalysis, Born2020POCOVID-net:POCUS, Roy2020DeepUltrasound, VanSloun2020LocalizingResults, Xue2021ModalityInformation, Gare2021DenseDetectionb}. Most of these methods rely on expert annotated data for learning, demanding scarce and expensive time from expert physicians and radiologists involved in the mitigation of the COVID-19 pandemic. This raises a need for label efficient learning techniques.



Monitoring patient severity and making prognostic predictions play a critical role in the allocation of limited medical resources. For this, several AI based patient severity scoring techniques have recently been proposed \cite{Roy2020DeepUltrasound, Xue2021ModalityInformation} which rely on video- and frame-based annotations. Labeling all of the individual frames in an ultrasound video clip is time-consuming and expensive though. Just labeling the ultrasound video clip is more suitable and treating the video clip severity label as the pseudo frame severity label for the corresponding frames of the video would be preferable. But doing so introduces label noise as not all the frames in a clip actually display the same severity sign. For instance, B-line artifact which is indicative of an unhealthy lung would not be consistently seen in all the frames of an unhealthy lung ultrasound clip, so not all the frames show the same level of disease state. We propose a contrastive learning strategy as a way to mitigate the label noise introduced by the use of such weak frame severity labels directly obtained from the corresponding video severity label. 


Contrastive learning has been used previously in the literature as semi- and self- supervised learning techniques \cite{Chen2020ARepresentations}, quite a few applications of it have already been presented in the medical domain \cite{ZhangCONTRASTIVETEXT, Wang2020ContrastiveClassification, Xue2021ModalityInformation}. Contrastive learning acts as a way to regularise feature embeddings to learn discriminative features that enforce intra-class features to have a greater overlap (or similarity) than inter-class features by using objective functions that operate on the cosine similarity of the feature embeddings. Many techniques apply contrastive learning for differentiating COVID-19, Healthy and other pneumonic diseases \cite{ZhangCONTRASTIVETEXT, Chen2020MomentumImages}. \citet{Chen2020MomentumImages} applied contrastive learning on CT scans as a few-shot COVID-19 diagnosis technique by bringing together the feature embedding of the same classes and pulling apart the feature embedding of different classes. Similarly, \citet{ZhangCONTRASTIVETEXT} applied contrastive learning on CT scans and paired text to enhance the network's domain invariance without using any expert annotation. \citet{Xue2021ModalityInformation} applied contrastive learning on the patient level feature embedding in an attempt to align features from 2 different modalities corresponding to LUS and clinical information, to predict the patient severity. The LUS feature embeddings are high level feature embeddings that are aggregated from frame level features to ultrasound zone level features. In addition to making the feature embedding of the two modalities align, they take care of preserving the patient severity discriminate features, by the introduction of novel additional loss components to the contrastive loss. Taking a cue from them, we also augment the contrastive loss with additional terms to retain the ultrasound severity discriminate features.  

We propose a weakly supervised training methodology by applying contrastive learning for the prediction of ultrasound video clip severity score, by making use of the noisy frame severity scores directly obtained from the corresponding video severity score. We show that the proposed contrastive learning setup is more robust to the weak frame severity label noise and thus generalizes better, compared to the cross-entropy loss based training.




\section{Methodology}

\subsubsection{Problem Statement}
Given an ultrasound B-mode grey image $I_g$, the task is to find a function $F \colon [ \, I_g] \, \to L$ that maps the image $I_g$ to ultrasound severity score labels $L \in \{0, 1, 2, 3\}$. Because the pleural line produces distinct artifacts (A-lines, B-lines) when scattering ultrasound based on the lung condition, the classification model should learn underlying mappings between the pleural line, artifacts, and pixel values, for making the predictions.  

\vspace{-2em}

\subsection{Data}

\begin{figure}[!tbp]
  \centering
  \resizebox{0.8\columnwidth}{!}{
  \begin{minipage}[b]{0.47\textwidth}
    \includegraphics[width=\textwidth]{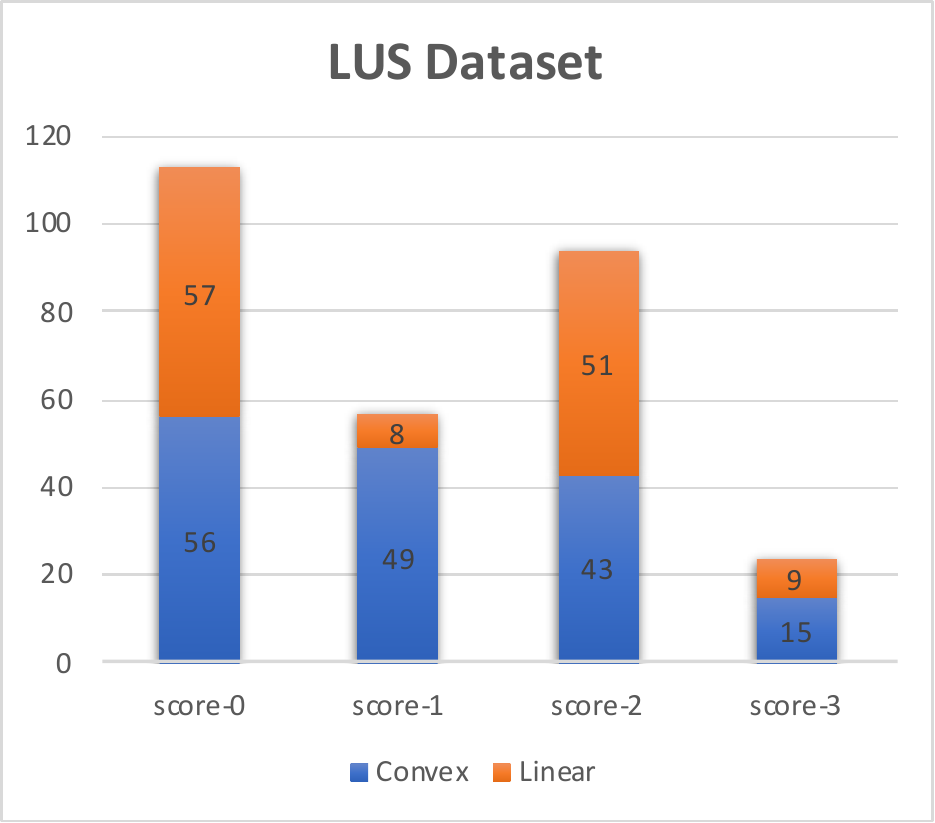}
    \caption{\small The distribution of ultrasound video clips into various severity scores and probes.}
  \label{fig:Lung-dataset}
  \end{minipage}
  \hfill
  \begin{minipage}[b]{0.47\textwidth}
    \includegraphics[width=\textwidth]{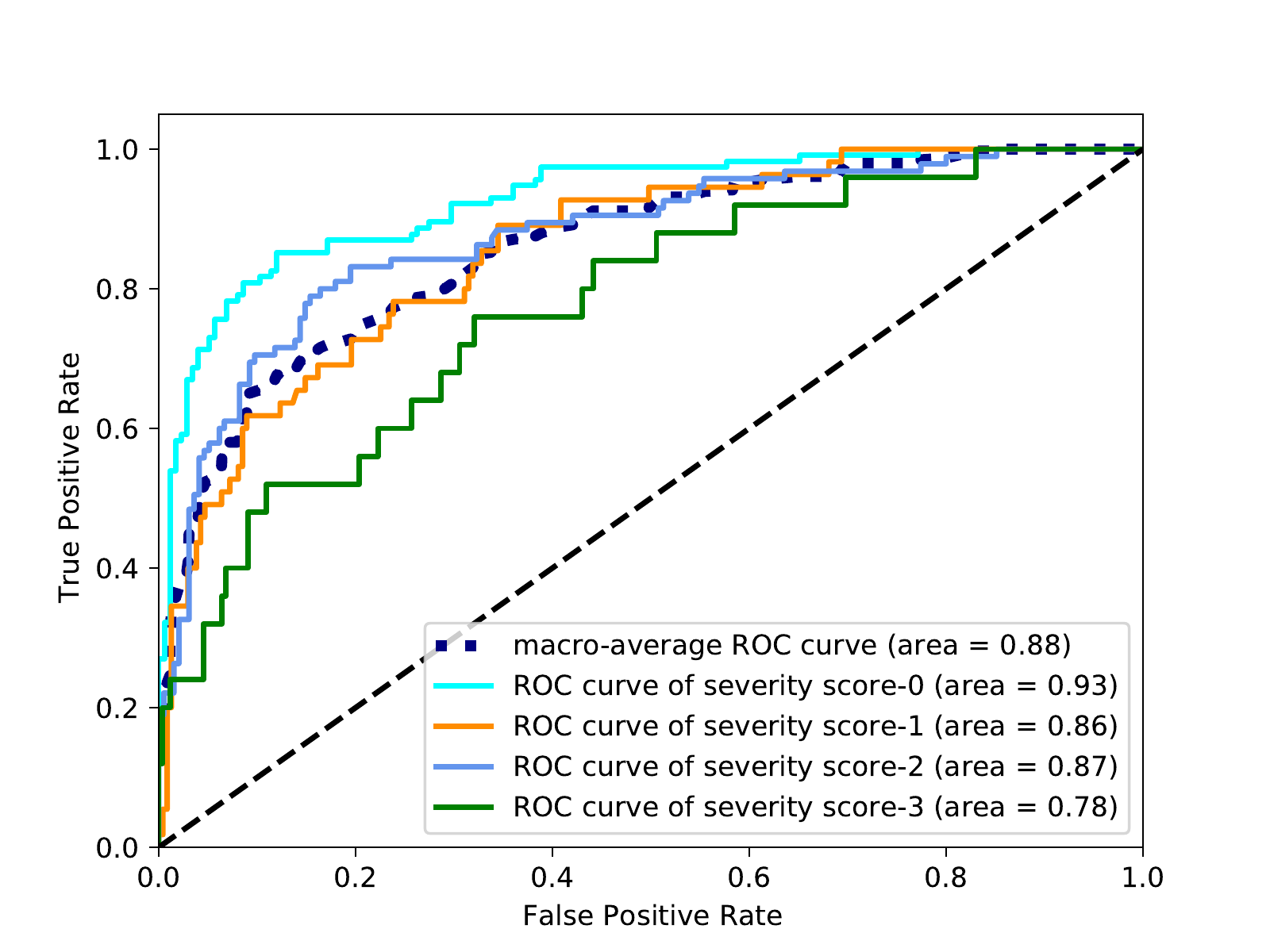}
    \caption{\small RoC plots of the contrastive learning trained model for the video-based severity scoring.}
   \label{fig:clr_roc_plot}
  \end{minipage}
  }
\end{figure}

We compiled a lung ultrasound dataset with linear and curvilinear videos sourced from the publicly usable subset of the POCOVID-Net dataset \cite{Born2020POCOVID-net:POCUS, Born2021AcceleratingAnalysis} (128 videos), as well as our own private dataset (160 videos). Our dataset consists of multiple ultrasound B-scans of left and right lung regions at depths ranging from 4cm to 6cm under different scan settings, obtained using a Sonosite X-Porte ultrasound machine. The combined dataset consists of ultrasound scans of healthy and COVID-19 patients, totaling 288 videos (113 Healthy and 175 COVID-19) resulting in about 50K images. \figureref{fig:Lung-dataset} shows the data distribution into the various ultrasound severity scores and probes.






We use the same 4-level ultrasound severity scoring scheme as defined in \cite{SimpleClinicalTrials.gov} and similarly used in \cite{Roy2020DeepUltrasound}. The score-0 indicates a normal lung with the presence of a continuous pleural line and horizontal A-line artifact. Scores 1 to 3 signify an abnormal lung, wherein score-1 indicates the presence of alterations in the pleural line with $\leq 5$ vertical B-line artifacts, score-2 has the presence of $> 5$ B-lines and score-3 signifies confounding B-lines with large consolidations. All the manual labeling was performed by individuals with at least a month of training from a pulmonary ultrasound specialist. Refer to \figureref{fig:gradcam_results} for sample images corresponding to the severity scores.



\subsubsection{Data Preprocessing}

We perform dataset upsampling to address the class imbalance for the training data, wherein we upsample all the minority class labeled data to get a balanced training dataset \cite{Rahman2013AddressingDatasets}. All the images are resized to 312x232 pixels using bilinear interpolation. Data augmentation is not applied.


\subsection{Training Strategy}
To access the ultrasound severity score of the video clips, we make use of the video labels as the noisy weak labels for the corresponding video frames. We augment the cross-entropy loss training objective for the classification task, using the contrastive learning objective in order to learn features that are robust to the frame-level label noise.

\subsubsection{Contrastive Learning Objective}


The proposed contrastive learning objective is inspired by \cite{Xue2021ModalityInformation}, wherein discriminative representations are learned using the contrastive loss consisting of three parts, which respectively cope with the intra-class alignment $\mathcal{L}^{IA}$, inter-class contrastive learning $\mathcal{L}^{CL}$, and contrastive continuity $\mathcal{L}^{CC}$. The intra-class alignment $\mathcal{L}^{IA}$ objective is to bring the feature embeddings of the same severity score closer, the inter-class contrastive learning $\mathcal{L}^{CL}$ objective is to differentiate the feature embeddings of different severity scores and the contrastive continuity $\mathcal{L}^{CC}$ ensure that the hierarchy among the severity scores is preserved. The proposed contrastive learning approach can be implemented by optimizing the following objective:

\begin{equation}
\label{eq:Lcon}
\mathcal{L}_{con} = \frac{1}{N} \sum^N_{i=1} [ \mathcal{L}_i^{IA} + \mathcal{L}_i^{CL} + \mathcal{L}_i^{CC}] 
\end{equation}

where,
\begin{equation}
\mathcal{L}_i^{IA} = 1 - sim(\mathbf{u}_i, \mathbf{u}_j) \quad \forall i, \exists j, |s_i - s_j| = 0 
\end{equation}

\begin{equation}
\mathcal{L}_i^{CL} = \sum_{s} sim(\mathbf{u}_k, \mathbf{u}_i)\quad \forall i, \exists k, |s_i - s_k| > 0    
\end{equation}

\begin{equation}
\mathcal{L}_i^{CC} = \sum_{s} max(sim(\mathbf{u}_m, \mathbf{u}_i) - sim(\mathbf{u}_n, \mathbf{u}_i), 0)
\end{equation}
$$\forall i, \exists m,n, |s_i - s_m| > 0, |s_i - s_n| > 0, |s_i - s_m| > |s_i - s_n|  $$ 

where, $N$ is the total number of frames, $sim(\mathbf{a}, \mathbf{b}) = \frac{\mathbf{a}^T \mathbf{b}}{\|\mathbf{a}\| \|\mathbf{b}\|}$ is the cosine similarity between vectors $\mathbf{a}$ and $\mathbf{b}$. $\mathbf{u}$ is the feature embeddings extracted after the global average pooling layer of the network, which is 2048-dimensional vector. $s$ is the ultrasound severity score of the corresponding frame feature $\mathbf{u}$. 

Unlike \cite{Xue2021ModalityInformation} which only relate the immediate severity levels, we explicitly relate all severity levels to enforce linear relationships in order to preserve the sequential nature of possible output choices (e.g. severity-1 is closer to severity-2 than severity-1 to severity-3) while simultaneously achieving the desired contrast in the loss. Our approach uniquely avoids the incorrect possibility of the model learning multi-dimensional distances among outputs, which could for example make severity-0 seem very close to severity-3 if the model incorrectly learned a cyclical order among the various severity levels. Prior systems do not take this ordinal relationship into account which can give rise to unnatural ordering. As can be observed in the confusion matrix shown in \figureref{fig:confusion_matrix}.


During training, for the input frame under consideration $i$, we randomly sample the frames $k,m,n$ from different video clips which have different severity scores than $i$ and randomly select frame $j$ corresponding to the same video clip as $i$ within a 10 frame window. 

\subsubsection{Overall Training Objective}
The overall training objective $\mathcal{L}_{overall}$ consists of the weighted combination of cross-entropy loss $\mathcal{L}_{ce}$ for classification error and contrastive learning loss $\mathcal{L}_{con}$ for feature regularization: 

\begin{equation}
\mathcal{L}_{overall} = \alpha \mathcal{L}_{ce} + (1 - \alpha) \mathcal{L}_{con} 
\end{equation}

where, the cross-entropy loss $\mathcal{L}_{ce} = \frac{1}{N} \sum_i - \mathbf{g}_i \log \mathbf{p}_i$, in which $N$ is the total number of frames, $\mathbf{g}_i$ is the ground truth one-hot severity score, $\mathbf{p}_i$ is the predicted probability scores from the last softmax layer of the network and the contrastive learning loss $\mathcal{L}_{con}$ is as defined in \equationref{eq:Lcon}. For all our experiments we set $\alpha$ as 0.5.





Using the frame predicted probability scores $\mathbf{p_i}$, we calculate the video's predicted probability scores $\mathbf{p}^v$ by taking the max severity-category score from all the corresponding video frame's predicted probability scores as:

\begin{equation}
\label{eq:video_prediction}
\mathbf{p}^v = softmax(\max_{i \in v} \mathbf{p_i[0]}, \max_{i \in v} \mathbf{p_i[1]}, \max_{i \in v} \mathbf{p_i[2]}, \max_{i \in v} \mathbf{p_i[3]})
\end{equation}

where, $\mathbf{p_i[0]}$, $\mathbf{p_i[1]}$, $\mathbf{p_i[2]}$, $\mathbf{p_i[3]}$, is severity category probability scores 0 to 3 respectively of frame $i$ belonging to video $v$. Using these video predicted probability scores $\mathbf{p}^v$ we evaluate the video-based severity scoring metrics of the model.

\subsubsection{Implementation}
The network is implemented with PyTorch and trained using the stochastic gradient descent algorithm \cite{Bottou2010Large-scaleDescent} with an Adam optimizer \cite{Kingma2015Adam:Optimization} set with an initial learning rate of $0.001$. The model is trained on an Nvidia Titan RTX GPU, with a batch size of 8 for 30 epochs for the classification task. The ReduceLRonPlateau learning-rate scheduler was used which reduces the learning rate by a factor (0.5) when the performance metric (accuracy) plateaus on the validation set. For the final evaluation, we pick the best model with the highest validation set accuracy to test on the held out test set.


\subsubsection{Metrics}
For the severity classification, we report accuracy, precision, recall, and F1 score \cite{Born2020POCOVID-net:POCUS, Roy2020DeepUltrasound}. The receiver operating characteristic (ROC) curve is also reported along with its area under the curve (AUC) metric \cite{Kim2020ChangesStudy}, wherein for the calculation of the metric the weighted average is taken, where the weights correspond to the support of each class and for the multi-label we consider the one-vs-all approach. \cite{Fawcett2006AnAnalysis}


\section{Experiment}

We train the ResNet-50 (RN50) \cite{He2016DeepRecognition} model, commonly used for classification and benchmarking methods using the proposed contrastive learning setup and compare its performance with the model trained only using the cross-entropy loss, in order to access the robustness achieved using the contrastive learning objective to the noisy weak frame severity score labels. We also compare the performance with the model trained using the original contrastive learning loss in \citet{Xue2021ModalityInformation} and a TSM  \cite{Lin2018TSM:Understanding} based video classification network similar to \cite{GareTheAI}, training details in Appendix-\ref{apn:tsm_training}. \emph{We conduct five independent runs, wherein each run we randomly split the videos into train, validation, and test sets with 70\%, 10\%, and 20\% split ratio respectively, by maintaining the same split ratio for all the individual severity scored clips and ensuring that all frames corresponding to a video remain in the same split.} The training set is upsampled to address the class imbalance \cite{Rahman2013AddressingDatasets}. We report the resulting metrics in form of mean and standard deviation over the five independent runs.

\vspace{-2em}

\begin{table*}[!ht]
\centering
\caption{Frame-based lung severity classification AUC of ROC, Accuracy, Precision, Recall, and F1 scores on lung dataset. Highest scores are shown in bold.}
\label{tab:frame_based_classification}
\resizebox{\textwidth}{!}{
\begin{tabular}{|c|c|c|c|c|c|c|c|}
 \hline
Method & AUC of ROC & accuracy & severity & precision & recall & F1-score\\
\hline



\hline

 \multirow{4}{*}{CE RN50} & \multirow{4}{*}{0.898 $\pm$ 0.016} & \multirow{4}{*}{0.693 $\pm$ 0.030} & score-0 &  0.872 $\pm$ 0.071 & 0.809 $\pm$ 0.037 & 0.836 $\pm$ 0.021 \\ 

 & & & score-1 &  0.529 $\pm$ 0.053 & 0.536 $\pm$ 0.195 & 0.517 $\pm$ 0.116 \\ 

 & & & score-2 &  0.763 $\pm$ 0.068 & 0.705 $\pm$ 0.089 & 0.727 $\pm$ 0.047 \\ 

 & & & score-3 &  0.167 $\pm$ 0.048 & 0.296 $\pm$ 0.067 & 0.212 $\pm$ 0.056 \\ 
\cline{4-7}
 & & & avg &  0.730 $\pm$ 0.038 & 0.693 $\pm$ 0.030 & 0.703 $\pm$ 0.035 \\ 
 
 \hline
 
 \multirow{4}{*}{proposed CL RN50} & \multirow{4}{*}{\bfseries{0.903 $\pm$ 0.022}} & \multirow{4}{*}{0.758 $\pm$ 0.042} & score-0 &  0.851 $\pm$ 0.039 & 0.886 $\pm$ 0.056 & 0.866 $\pm$ 0.016 \\ 

 & & & score-1 &  0.610 $\pm$ 0.131 & 0.612 $\pm$ 0.212 & 0.599 $\pm$ 0.156 \\ 

 & & & score-2 &  0.775 $\pm$ 0.070 & 0.771 $\pm$ 0.040 & 0.771 $\pm$ 0.041 \\ 

 & & & score-3 &  0.373 $\pm$ 0.168 & 0.223 $\pm$ 0.099 & 0.264 $\pm$ 0.100 \\ 
\cline{4-7}
 & & & avg &  0.752 $\pm$ 0.048 & 0.758 $\pm$ 0.042 & 0.748 $\pm$ 0.044 \\ 

\hline

 \multirow{4}{*}{original CL RN50} & \multirow{4}{*}{0.899 $\pm$ 0.020} & \multirow{4}{*}{\bfseries{0.759 $\pm$ 0.041}} & score-0 &  0.855 $\pm$ 0.056 & 0.915 $\pm$ 0.024 & 0.883 $\pm$ 0.033 \\ 

 & & & score-1 &  0.620 $\pm$ 0.060 & 0.555 $\pm$ 0.081 & 0.583 $\pm$ 0.065 \\ 

 & & & score-2 &  0.764 $\pm$ 0.021 & 0.761 $\pm$ 0.076 & 0.760 $\pm$ 0.038 \\ 

 & & & score-3 &  0.429 $\pm$ 0.294 & 0.295 $\pm$ 0.142 & 0.318 $\pm$ 0.171 \\ 
\cline{4-7}
 & & & avg &  \bfseries{0.754 $\pm$ 0.046} & \bfseries{0.759 $\pm$ 0.041} & \bfseries{0.752 $\pm$ 0.041} \\

\hline
\end{tabular}
}
\end{table*}


\begin{table}[!ht]
\centering
\caption{Video-based lung severity classification AUC of ROC, Accuracy, Precision, Recall, and F1 scores on lung dataset. Highest scores are shown in bold.}
\label{tab:video_based_classification}
\resizebox{\textwidth}{!}{
\begin{tabular}{|c|c|c|c|c|c|c|c|}
 \hline
Method & AUC of ROC & accuracy & severity & precision & recall & F1-score\\
\hline



 \multirow{4}{*}{CE RN50} & \multirow{4}{*}{0.842 $\pm$ 0.027} & \multirow{4}{*}{0.655 $\pm$ 0.055} & score-0  &  0.851 $\pm$ 0.083 & 0.739 $\pm$ 0.027 & 0.788 $\pm$ 0.036 \\ 

 & & & score-1 &  0.523 $\pm$ 0.058 & 0.527 $\pm$ 0.156 & 0.516 $\pm$ 0.098 \\ 

 & & & score-2 &  0.751 $\pm$ 0.088 & 0.684 $\pm$ 0.120 & 0.708 $\pm$ 0.077 \\ 

 & & & score-3 &  0.243 $\pm$ 0.095 & 0.440 $\pm$ 0.150 & 0.312 $\pm$ 0.116 \\ 
\cline{4-7}
 & & & avg &  0.704 $\pm$ 0.053 & 0.655 $\pm$ 0.055 & 0.669 $\pm$ 0.055 \\ \hline

 \multirow{4}{*}{proposed CL RN50} & \multirow{4}{*}{0.867 $\pm$ 0.020} & \multirow{4}{*}{\bfseries{0.734 $\pm$ 0.065}} & score-0 &   0.832 $\pm$ 0.051 & 0.843 $\pm$ 0.071 & 0.835 $\pm$ 0.044 \\ 

 & & & score-1 &  0.630 $\pm$ 0.162 & 0.636 $\pm$ 0.199 & 0.621 $\pm$ 0.154 \\ 

 & & & score-2 &  0.761 $\pm$ 0.095 & 0.768 $\pm$ 0.071 & 0.761 $\pm$ 0.060 \\ 

 & & & score-3 &  0.457 $\pm$ 0.290 & 0.320 $\pm$ 0.160 & 0.364 $\pm$ 0.201 \\ 
\cline{4-7}
 & & & avg & 0.738 $\pm$ 0.068  & \bfseries{0.734 $\pm$ 0.065} & \bfseries{0.730 $\pm$ 0.064} \\ \hline

 \multirow{4}{*}{original CL RN50} & \multirow{4}{*}{0.879 $\pm$ 0.026} & \multirow{4}{*}{0.731 $\pm$ 0.036} & score-0  &  0.819 $\pm$ 0.077 & 0.861 $\pm$ 0.017 & 0.837 $\pm$ 0.040 \\ 

 & & & score-1 &  0.639 $\pm$ 0.026 & 0.582 $\pm$ 0.093 & 0.606 $\pm$ 0.058 \\ 

 & & & score-2 &  0.763 $\pm$ 0.048 & 0.747 $\pm$ 0.117 & 0.747 $\pm$ 0.051 \\ 

 & & & score-3 &  0.503 $\pm$ 0.261 & 0.400 $\pm$ 0.219 & 0.396 $\pm$ 0.130 \\ 
\cline{4-7}
 & & & avg &  0.739 $\pm$ 0.045 & 0.731 $\pm$ 0.036 & 0.726 $\pm$ 0.036 \\ \hline

 \multirow{4}{*}{CE TSM} & \multirow{4}{*}{\bfseries{0.897 $\pm$ 0.025}} & \multirow{4}{*}{0.710 $\pm$ 0.060} & score-0 &  0.911 $\pm$ 0.059 & 0.730 $\pm$ 0.139 & 0.801 $\pm$ 0.082 \\ 

 & & & score-1 &  0.604 $\pm$ 0.081 & 0.764 $\pm$ 0.109 & 0.672 $\pm$ 0.079 \\ 

 & & & score-2 &  0.745 $\pm$ 0.085 & 0.768 $\pm$ 0.026 & 0.755 $\pm$ 0.056 \\ 

 & & & score-3 &  0.276 $\pm$ 0.097 & 0.280 $\pm$ 0.098 & 0.270 $\pm$ 0.089 \\ 
\cline{4-7}
 & & & avg &  \bfseries{0.744 $\pm$ 0.036} & 0.710 $\pm$ 0.060 & 0.716 $\pm$ 0.054 \\

\hline
\end{tabular}
}
\end{table}

\vspace{-2em}

\section{Results and Discussions}		

\tableref{tab:frame_based_classification} shows the mean and standard deviation of the frame-based severity scoring metrics, obtained by evaluating on the held-out test set using the models from the five independent runs. We observe that the contrastive learning (CL) based trained models preform better than the cross-entropy (CE) trained model, wherein the original and the proposed contrastive learning loss have similar scores with the original loss performing slightly better.  

\begin{figure}[!tbp]
  \centering
  \begin{minipage}[b]{0.47\textwidth}
    \includegraphics[width=0.8\textwidth]{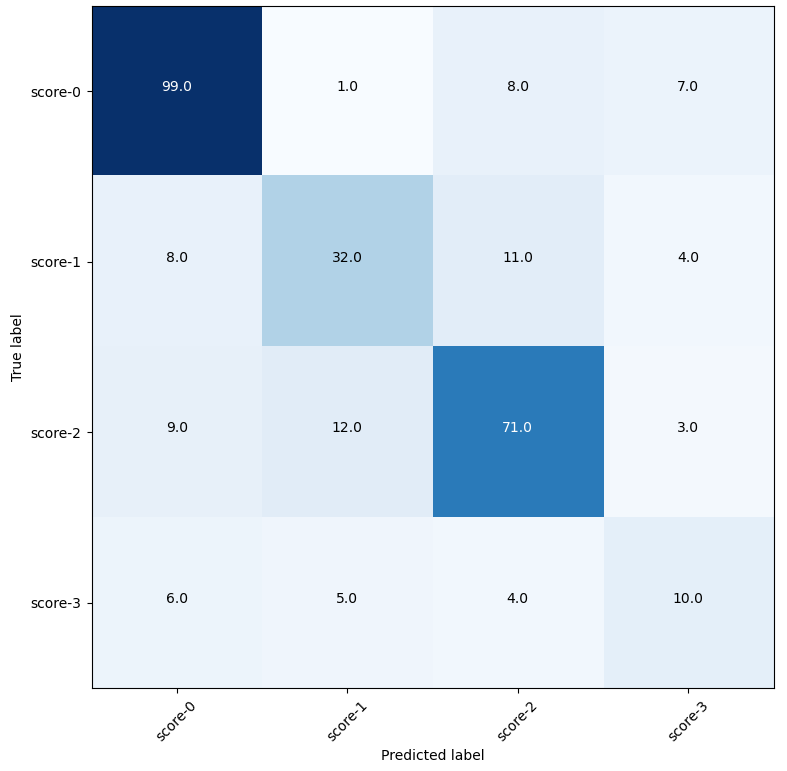}
  \end{minipage}
  \hfill
  \begin{minipage}[b]{0.47\textwidth}
    \includegraphics[width=0.8\textwidth]{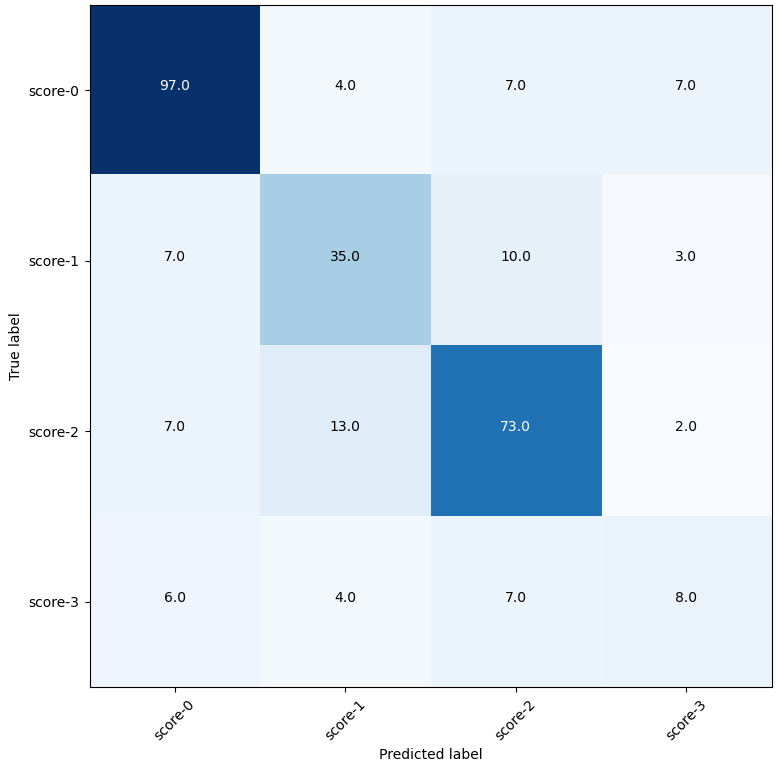}
  \end{minipage}
  \label{fig:confusion_matrix}
  \caption{\small Confusion matrix of the contrastive learning loss original (left) vs proposed (right). Our proposed loss is confused between immediate severity scores which is reasonable and is less confused between non-immediate severity scores compared to the original loss.
  }
\end{figure}

We calculate the video-based severity scoring metrics of the models by calculating the video predicted probability score $\mathbf{p}^v$ obtained by taking the max severity-category score from all the corresponding video frame's predicted probability scores $\mathbf{p}$, as defined in \equationref{eq:video_prediction}. \tableref{tab:video_based_classification} shows the mean and standard deviation of the video-based severity scoring metrics, obtained by evaluating on the held out test set using the models from the five independent runs. We again observe that the contrastive learning (CL) based trained models preform better than the cross-entropy (CE) trained model and has comparable performance with the video based TSM model. With our proposed loss function achieving the highest accuracy, recall, and F1-score. The macro average and individual severity score's RoC plots of the CL trained model using the proposed loss for video-based prediction is shown in \figureref{fig:clr_roc_plot}. The lower performance on severity score-3 compared to other scores could be due to the limited number of training data for severity score-3. \figureref{fig:confusion_matrix} shows the confusion matrix of both the contrastive loss trained models on the combined 5 runs.


On comparing the model's scoring metrics on the held out test set with the validation (val) set used for hyperparameter optimization (see \tableref{tab:test_vs_val}), we see that though the CE trained model achieved higher accuracy and F1-score (avg) on the validation set compared to our CL trained model, it was outperformed on the held out test set by the CL trained model. This suggests that the CL trained model generalized better to the unseen data, which is indicative of robust features learned using the contrastive loss. 


We visualize the model's layer-2 Grad-CAM \cite{Selvaraju2016Grad-CAM:Localization} and show the mean Grad-CAM image corresponding to the four severity scores taken over the entire test set ($\sim10$K images) for the best run in \figureref{fig:gradcam_results}. We also shown Grad-CAM on four randomly selected images for which our CL trained model appeared to be looking at the correct locations (pleural line and A-line \& B-line artifacts), whereas CE trained model was basing its predictions on non-lung tissue. For these four test images, the CL model correctly predicted the severity scores, whereas the CE model got all predictions wrong. Which suggests that the contrastive learning objective lead to learning better discriminative features.

\vspace{-2em}
  
\begin{table*}[!ht]
\centering
\caption{Performance comparison of frame-based score prediction on Test and Val dataset.}
\label{tab:test_vs_val}
\resizebox{0.8\textwidth}{!}{
\begin{tabular}{|c|c|c|c|c|}
 \hline
Dataset & Method & AUC of ROC & accuracy & F1-score\\
\hline

\multirow{2}{*}{Test set}  & CE RN50 & 0.898 $\pm$ 0.016 & 0.693 $\pm$ 0.030 & 0.703 $\pm$ 0.035 \\
  & CL RN50 & \bfseries{0.903 $\pm$ 0.022} & \bfseries{0.758 $\pm$ 0.042} & \bfseries{0.748 $\pm$ 0.043} \\

\hline

\multirow{2}{*}{Val set}  & CE RN50 & 0.837 $\pm$ 0.074 & \bfseries{0.689 $\pm$ 0.094} & \bfseries{0.685 $\pm$ 0.093} \\
  & CL RN50 & \bfseries{0.839 $\pm$ 0.048} & 0.652 $\pm$ 0.069 & 0.633 $\pm$ 0.091 \\

\hline
\end{tabular}
}
\end{table*}

\newlength{\width}
\setlength{\width}{0.55 in}
\newlength{\height}
\setlength{\height}{0.45 in}

\begin{figure}[!ht] 
\centering

\setlength{\tabcolsep}{1pt} 
\def\arraystretch{0.5} 

\newcolumntype{C}{>{\centering\arraybackslash}m{\width}<{}}
\newcolumntype{F}{>{\centering\arraybackslash}m{0.3\width}<{}}

\resizebox*{\columnwidth}{0.35\textheight}{
\begin{tabular}{FF CC CC}

&
&
\tiny score-0 &
\tiny score-1 &
\tiny score-2 &
\tiny score-3 \\

\rotatebox[origin=c]{90}{\centering \tiny grey} & &
\includegraphics[height = \height, width = \width]{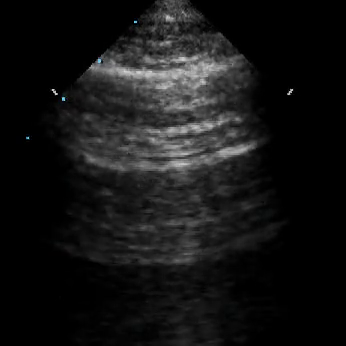} &
\includegraphics[height = \height, width = \width]{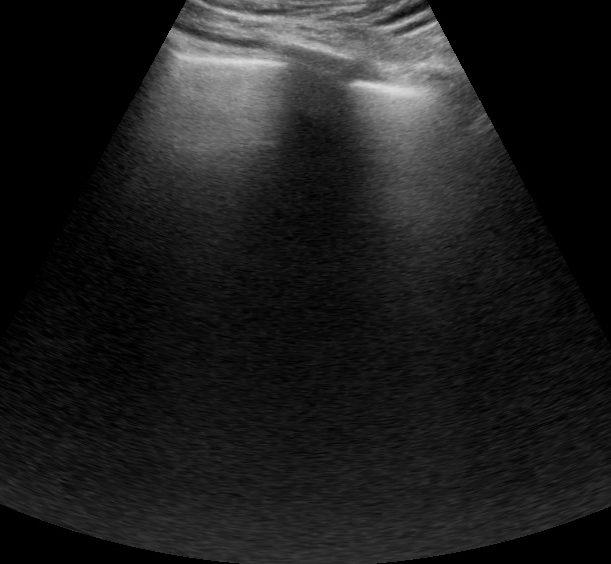} &
\includegraphics[height = \height, width = \width]{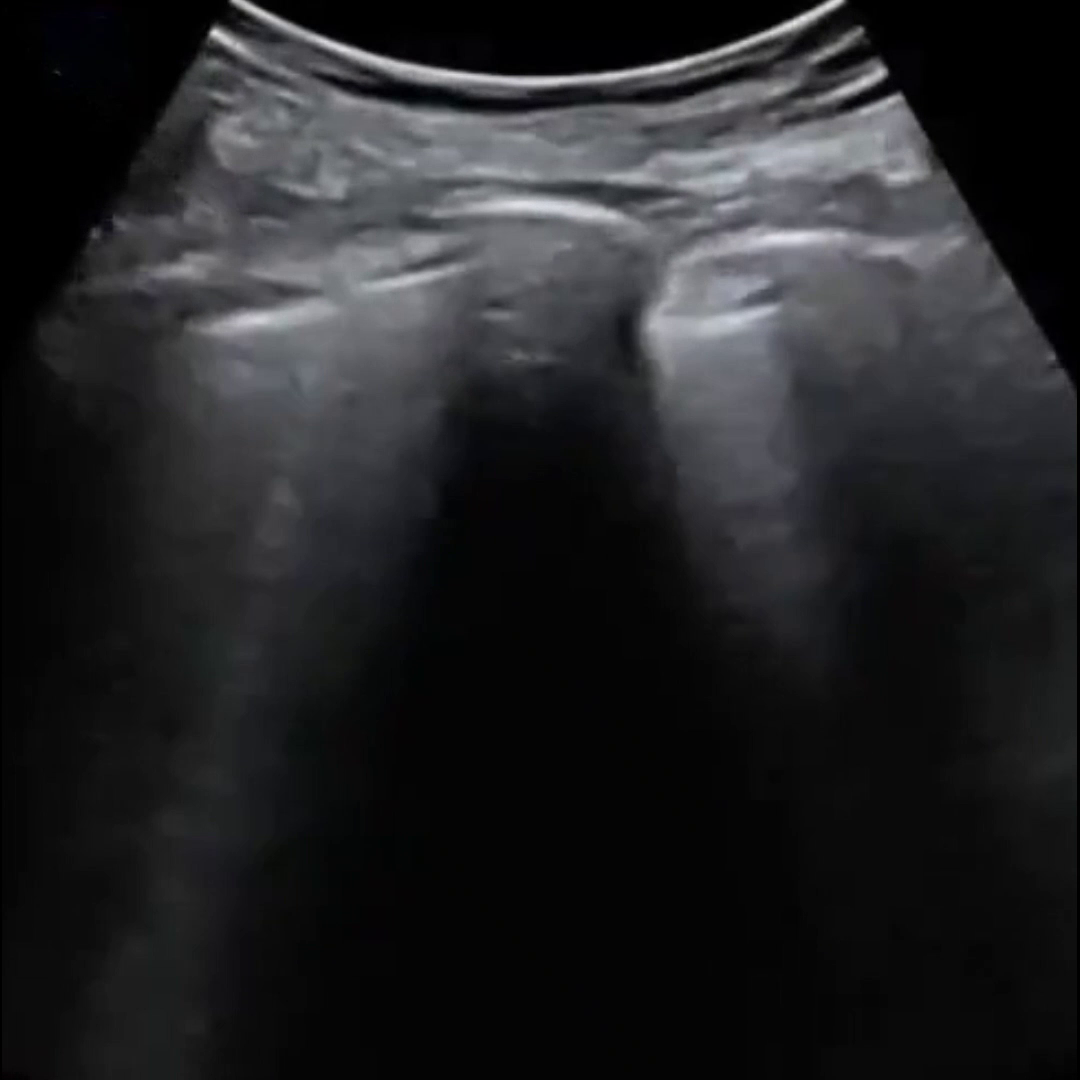} &
\includegraphics[height = \height, width = \width]{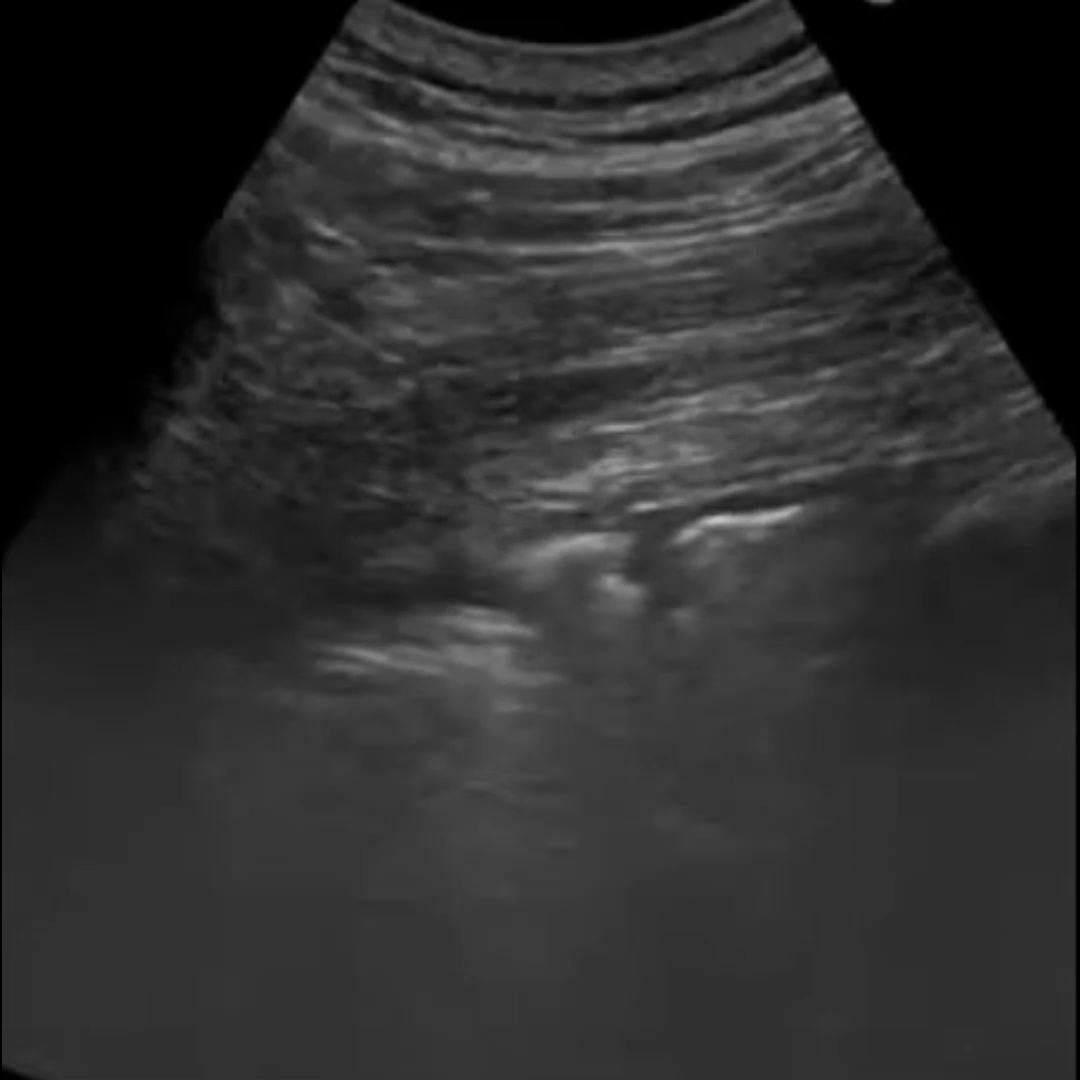} \\

\multirow{2}{*}{\raisebox{0.4cm}{\rotatebox[origin=c]{90}{\parbox{1.5\height}{\centering \tiny random sample}}}} &
\raisebox{0.4cm}{\rotatebox[origin=c]{90}{\parbox{\height}{\centering \tiny CE RN50}}} &
\includegraphics[height = \height, width = \width]{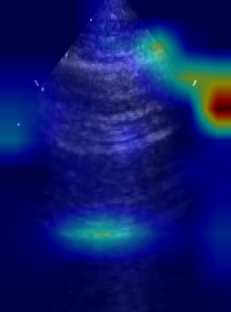} &
\includegraphics[height = \height, width = \width]{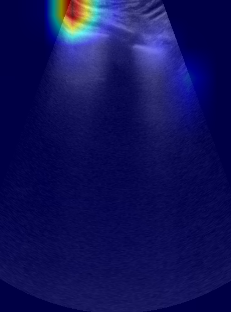} &
\includegraphics[height = \height, width = \width]{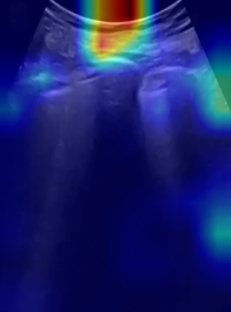} &
\includegraphics[height = \height, width = \width]{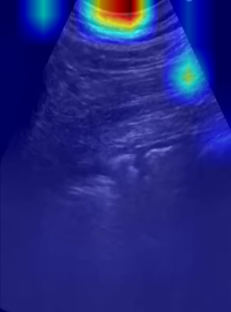} \\ [-0ex]

&
\raisebox{0.4cm}{\rotatebox[origin=c]{90}{\parbox{\height}{\centering \tiny CL RN50}}} &
\includegraphics[height = \height, width = \width]{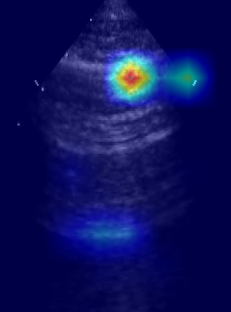} &
\includegraphics[height = \height, width = \width]{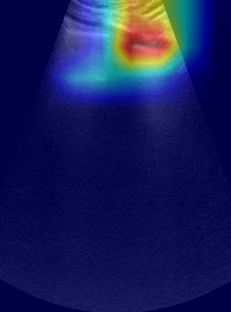} &
\includegraphics[height = \height, width = \width]{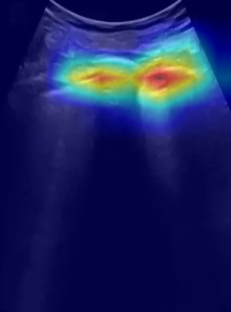} &
\includegraphics[height = \height, width = \width]{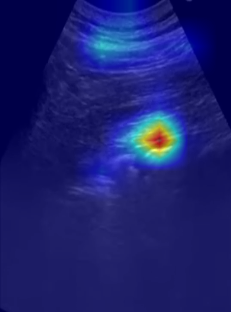} \\

\multirow{2}{*}{\raisebox{0.4cm}{\rotatebox[origin=c]{90}{\parbox{1.5\height}{\centering \tiny mean over testset}}}} &
\raisebox{0.4cm}{\rotatebox[origin=c]{90}{\parbox{\height}{\centering \tiny CE RN50}}} &
\includegraphics[height = \height, width = \width]{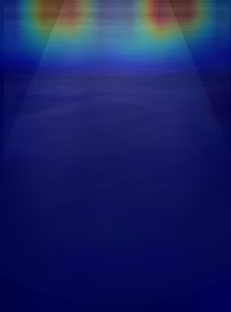} &
\includegraphics[height = \height, width = \width]{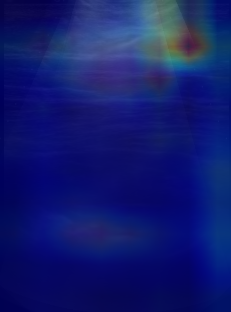} &
\includegraphics[height = \height, width = \width]{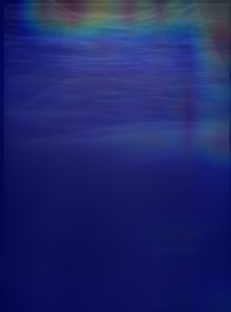} &
\includegraphics[height = \height, width = \width]{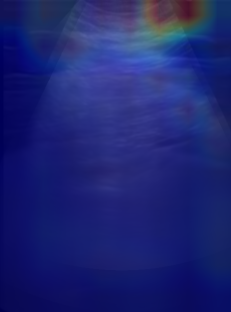} \\ [-0ex]

&
\raisebox{0.4cm}{\rotatebox[origin=c]{90}{\parbox{\height}{\centering \tiny CL RN50}}} &
\includegraphics[height = \height, width = \width]{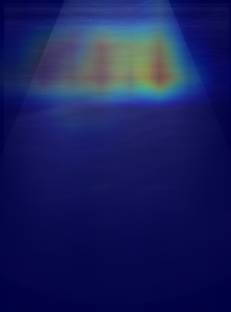} &
\includegraphics[height = \height, width = \width]{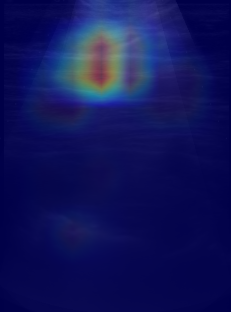} &
\includegraphics[height = \height, width = \width]{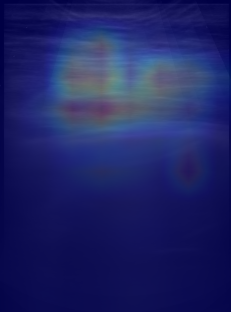} &
\includegraphics[height = \height, width = \width]{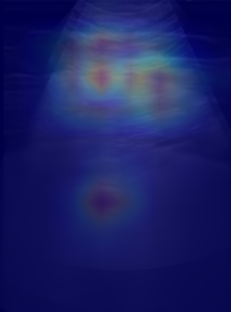} \\

\end{tabular}

}

\caption{
\small Grad-CAM \cite{Selvaraju2016Grad-CAM:Localization} visualization of the layer-2 of cross-entropy (CE) and contrastive learning (CL) trained model on the four severity score test images (B-mode grey). We observe that CL trained model bases the predictions predominantly on the pleural line and A-line \& B-line artifacts, whereas the CE trained model predominantly bases the predictions on the subcutaneous tissues above the pleural line.}
\label{fig:gradcam_results}
\end{figure}

\vspace{-2em}

\section{Conclusion}

We demonstrated a weakly supervised method for scoring the COVID-19 lung ultrasound scan clips, using our proposed contrastive learning objective. Which treats video-based severity labels as frame-based severity labels thus reducing labeling cost. While these frame labels are noisy, we demonstrated that the contrastive learning objective is robust to such label noise compared to the cross-entropy learning objective. We showed that the frame based model trained using the proposed contrastive learning loss achieves comparable performance to a video based TSM model. 
\midlacknowledgments{
This present work was sponsored in part by US Army Medical contracts W81XWH-19-C0083 and W81XWH-19-C0101. This work used the Extreme Science and Engineering Discovery Environment (XSEDE), which is supported by National Science Foundation grant number ACI-1548562. Specifically, it used the Bridges system, which is supported by NSF award number ACI-1445606, at the Pittsburgh Supercomputing Center (PSC). We would also like to thank our collaborators at the Carnegie Mellon University (CMU), Louisiana State University (LUS), and University of Pittsburgh (Upitt). 
We are pursuing intellectual-property protection. Galeotti serves on the advisory board of Activ Surgical, Inc. He and Rodriguez are involved in the startup Elio AI, Inc.
}

\bibliography{references}

\appendix

\section{TSM model Training Strategy}
\label{apn:tsm_training}
We follow the same setup of \cite{GareTheAI} for training a TSM network \cite{Lin2018TSM:Understanding} with ResNet-18 (RN18) \cite{He2016DeepRecognition} backbone and bi-directional residual shift with $1/8$ channels shifted in both directions.
The model is fed input clips of 16 frames wide (224x224 pixels) sampled using the same strategy as in \citet{GareTheAI}. For testing, 3 sequential sample clips per video are evaluated which are used to get the corresponding video predicted probability scores $\mathbf{p^v}$, as defined in \equationref{eq:video_prediction}.
The model is trained for 30 epochs using cross-entropy loss. For fair comparison with the frame based models no augmentation is used. 

\section{Comparison with other existing work}

We compare our video-based scoring with prior method reported scores in the literature \cite{Roy2020DeepUltrasound, Xue2021ModalityInformation} in \tableref{tab:other_methods}. We see that our method achieves higher scores, though noting that these scores are obtained on different datasets. 

\begin{table*}[!ht]
\centering
\caption{Performance comparison of the video-based score prediction with other existing work. Scores are obtained on different dataset.}
\label{tab:other_methods}
\resizebox{\textwidth}{!}{
\begin{tabular}{|c|c|c|c|c|c|c|}
 \hline
Method & AUC of ROC & accuracy & precision & recall & F1-score\\
\hline

\cite{Roy2020DeepUltrasound}  & - & - & 0.70 & 0.60 & 0.61 \\

\cite{Xue2021ModalityInformation} & - & 0.5660 & 0.5648 & 0.5630 & 0.5639 \\

ours  & 0.867 & 0.734 & 0.738 & 0.734 & 0.730 \\

\hline
\end{tabular}
}
\end{table*}

\end{document}